\documentclass[11pt]{article}

\usepackage{arxiv}

\usepackage[utf8]{inputenc} 
\usepackage[T1]{fontenc}    
\usepackage{hyperref}       
\usepackage{url}            
\usepackage{booktabs}       
\usepackage{amssymb,amsmath,mathtools,amsfonts}       
\usepackage{bm}
\usepackage{xfrac}       
\usepackage{microtype}      
\usepackage{doi}
\usepackage[table,xcdraw]{xcolor}
\usepackage{graphicx}
\usepackage[algoruled,noline,noend]{algorithm2e}
\usepackage{pgfplots}
\pgfplotsset{compat=1.18}

\usepackage{subcaption}
\usepackage[%
    backend=biber,%
    style=ieee,%
    mincitenames=1,%
    maxcitenames=1,%
]{biblatex}
\renewcommand{\vec}[1]{\bm{#1}}

\title{Extending the Lattice Boltzmann Method to Non-linear Solid Mechanics}

\date{2 February 2025} 					

\newcommand{\orcid}[2]{\href{https://orcid.org/#1}{\includegraphics[width=1.6ex]{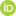}\hspace{1mm}{#2}}}

\author{ %
    \orcid{0000-0003-4819-2198}{Henning Müller}\thanks{Corresponding author} \\
	Technische Universität Darmstadt\\
	Institut für Mechanik\\
	\texttt{henning.mueller@tu-darmstadt.de} \\
    \And
	\orcid{0000-0003-2189-945X}{Erik Faust} \\
	RPTU Kaiserslautern-Landau\\
	Lehrstuhl für Technische Mechanik\\
	\texttt{erik.faust@mv.rptu.de} \\
    \And
	\orcid{0000-0002-0839-6519}{Alexander Schlüter} \\
	Technische Universität Darmstadt\\
	Institut für Mechanik\\
	\texttt{alexander.schlueter@tu-darmstadt.de} \\
	\And
	{Ralf Müller} \\
	Technische Universität Darmstadt\\
	Institut für Mechanik\\
	\texttt{ralf.mueller@mechanik.tu-darmstadt.de} \\
}

\renewcommand{\headeright}{}
\renewcommand{\undertitle}{}
\renewcommand{\shorttitle}{Extending the LBM to Non-linear Solid Mechanics}

\hypersetup{
pdftitle={Extending the LBM to Non-linear Solid Mechanics},
pdfsubject={cs.CE, ma.NA},
pdfauthor={Henning Müller, Erik Faust, Alexander Schlüter, Ralf Müller},
pdfkeywords={Computational Solid Mechanics, Lattice Boltzmann, Non-linear Solid Mechanics, Elastodynamics, Finite Strains},
}

\begin{document}

\maketitle

\begin{abstract}
    This work outlines a Lattice Boltzmann Method (LBM) for geometrically and constitutively non-linear solid mechanics to simulate large deformations under dynamic loading conditions.
    The method utilizes the moment chain approach, where the non-linear constitutive law is incorporated via a forcing term.
    Stress and deformation measures are expressed in the reference configuration.
    Finite difference schemes are employed for gradient and divergence computations, and Neumann- and Dirichlet-type boundary conditions are introduced.

    Numerical studies are performed to assess the proposed method and illustrate its capabilities.
    Benchmark tests for weakly dynamic uniaxial tension and simple shear across a range of Poisson's ratios demonstrate the feasibility of the scheme and serve as validation of the implementation.
    Furthermore, a dynamic test case involving the propagation of bending waves in a cantilever beam highlights the potential of the method to model complex dynamic phenomena.
\end{abstract}


\section{Introduction}

Lattice Boltzmann methods (LBMs) are established solvers in computational fluid dynamics.
They are computationally efficient, work with a simple discretisation~\cite{kruger_lattice_2017} and are quite versatile, allowing them to be used for multiphysical simulations.

In recent years, efforts have been undertaken to develop Lattice Boltzmann (LB) schemes for solid mechanics.
\textcite{murthy_lattice_2018} and \textcite{escande_lattice_2020} introduced a moment chain algorithm for linear elastodynamics, which is close to the LB formulation for fluids.
This approach has been extended for Neumann- and Dirichlet-type boundary conditions by \textcite{faust_dirichlet_2024} and subsequently been used for dynamic simulations, such as transient models~\cite{faust_modelling_2023} and crack propagation~\cite{muller_dynamic_2023}.
Other recent endeavours~\cite{maquart_toward_2022,boolakee_new_2023,boolakee_dirichlet_2023} cover stationary systems.
So far, all approaches have assumed linear elasticity and small deformations, limiting their scope to the regime where non-linear effects and large deformations are negligible.

This work proposes an extension of the moment chain LBM to the non-linear regime, both geometrically, with large deformations, and constitutively, with hyperelastic material laws.
The moment chain is formulated in the reference configuration.
This is common for solid mechanics and also simplifies the implementation.
No changes to the computational domain, i.e.\ the lattice representing the material domain, are necessary even for large deformations.
This work focusses on the feasibility of the proposed method.

Sect.~\ref{sec:balance} recapitulates the LB method with an emphasis on solving systems of balance equations.
Next, in Sect.~\ref{sec:solidlbm}, the moment chain for solid mechanics and the resulting LB algorithm are introduced. This includes boundary conditions.
Numerical examples for St.\ Venant-Kirchhoff-- and neo-Hooke--type materials are described in Sect.~\ref{sec:num}.
Benchmarks for uniaxial tension and simple shear are treated to validate the algorithm. 
An example of propagation of bending waves in a cantilever beam shows the possibility to simulate larger and more complex dynamical systems.
Sect.~\ref{sec:results} comprises a discussion of the results, followed by a summary in Sect.~\ref{sec:summary}.

\section{Lattice Boltzmann for balance equations}
\label{sec:balance}

This section briefly discusses the central steps of the LB algorithm and the connection to balance laws.
The interested reader is referred to textbooks, such as \textcite{kruger_lattice_2017,succi_lattice_2018}, which cover the LBM in detail.
Any definitions relating to the specific physical system under consideration will be given in Sect.~\ref{sec:solidlbm} in the context of solid mechanics.

LB schemes utilize a statistical approach, originally based on kinetic theory.
The state of the system is described in terms of distribution functions $f(\vec{x}, \vec{\xi}, t)$ in spatial coordinates $\vec{x}$ and velocities $\vec{\xi}$.
The time evolution is then given by the Boltzmann equation
\begin{gather}
    \frac{\partial f}{\partial t} + \vec{\xi} \cdot \frac{\partial f}{\partial \vec{x}} + \frac{\vec{F}}{\rho} \cdot \frac{\partial f}{\partial \vec{\xi}} = \Omega (f),
    \label{eq:balance:boltzmann}
\end{gather}
where $\vec{F}$ describes an external force and $\rho$ is the mass density.
$\Omega$ is the collision operator that captures particle collisions as described by kinetic theory.
The simplest formulation for $\Omega$ is the single-time linearized BGK operator\cite{bhatnagar_model_1954,welander_temperature_1954}
\begin{gather}
    \Omega^\text{BGK} = \frac{1}{\tau} \left( f - f^\text{eq} \right),
    \label{eq:balance:bgk}
\end{gather}
which describes a relaxation of the distribution $f$ toward the local equilibrium $f^\text{eq}$ with the relaxation time $\tau$.
The equilibrium distribution $f^\text{eq}$ must be defined according to the specific physical system to be modelled.

For discretization, the physical space is represented by a regular lattice with spacing $\Delta X$ and the velocity space is discretized by a fixed set of velocities.
Here, the D2Q9-scheme is used, which is a set of 9 velocities in 2 dimensions that effectively connect a lattice node to its neighbors by defining the transport of distribution functions.
The lattice velocity $c = \sfrac{\Delta X}{\Delta t}$ ensures the exchange of information between these points in one time step and couples the spatial with the temporal discretisation.
The set of velocities is given in Tab.~\ref{tab:balance:scheme} in terms of $c$, which will be defined for the case of solids later.
Also shown are weights for every lattice direction, which originate from numerical quadratures~\cite{kruger_lattice_2017} and are needed to project quantities on the lattice directions.
These will be defined in Sect.~\ref{sec:solidlbm}, when definitions are given, e.g.\ for $f^\text{eq}$.

\begin{table}[bt]
    \centering
    \caption{%
        The velocities and weights for index $q$ in the \textit{D2Q9}-lattice scheme
    }
    \label{tab:balance:scheme}
    \begin{tabular}{@{} c | c c c c c c c c c @{}}
        \toprule
        $q$
            & 0 & 1 & 2 & 3 & 4 & 5 & 6 & 7 & 8 \\
        \midrule
        $\boldsymbol{c}_q$
            & $\begin{pmatrix} 0 \\ 0\end{pmatrix}$   
            & $\begin{pmatrix} c \\ 0\end{pmatrix}$   
            & $\begin{pmatrix} 0 \\ c\end{pmatrix}$   
            & $\begin{pmatrix} -c \\ 0\end{pmatrix}$  
            & $\begin{pmatrix} 0 \\ -c\end{pmatrix}$  
            & $\begin{pmatrix} c \\ c\end{pmatrix}$   
            & $\begin{pmatrix} -c \\ c\end{pmatrix}$  
            & $\begin{pmatrix} -c \\ -c\end{pmatrix}$ 
            & $\begin{pmatrix} c \\ -c\end{pmatrix}$  
            \\
        \midrule
        $w_q$
            & $\dfrac{4}{9}$     
            & $\dfrac{1}{9}$     
            & $\dfrac{1}{9}$     
            & $\dfrac{1}{9}$     
            & $\dfrac{1}{9}$     
            & $\dfrac{1}{36}$    
            & $\dfrac{1}{36}$    
            & $\dfrac{1}{36}$    
            & $\dfrac{1}{36}$    
            \\
        \bottomrule
    \end{tabular}
\end{table}

This discretization leads to the Lattice Boltzman equation (LBE)
\begin{gather}
    f_q (\vec{x} + \vec{c}_q \Delta t, t + \Delta t) = f_q (\vec{x}, t) + \Omega_q \Delta t.
    \label{eq:balance:lbe}
\end{gather}

In the following, the BGK-collision operator Eq. \eqref{eq:balance:bgk} is used and a term $\psi$, taking contributions of foces into account, is added.
Furthermore, a split into a collision and a streaming step is undertaken.
In the collision step, the BGK-LBE
\begin{gather}
    f_q^\text{coll} = f_q - \frac{\Delta t}{\tau} (f_q - f_q^\text{eq}) + \Delta t \left( 1 - \frac{\Delta t}{2 \tau} \psi_q \right)
    \label{eq:balabce:lbe-coll}
\end{gather}
is evaluated.
This is an explicit and local operation, restricted to individual lattice nodes, and can thus easily be executed in parallel.
Afterwards, theses post-collision distributions $f_q^\text{coll}$ are streamed to the neighbours, i.e.
\begin{gather}
    f_q (\vec{x} + \vec{c}_q \Delta t, t + \Delta t) = f_q^\text{coll}.
\end{gather}
The split into collision and streaming brings great computational advantages in terms of efficiency.

\subsection{The moment-chain approach}

Take a continuum mechanical balance equation of the general format
\begin{gather}
    \mathrm{d}_t \mathcal{M}^{(N)}_{[\alpha]} + \mathrm{d}_\beta \mathcal{M}^{(N+1)}_{[\alpha]\beta} = S^{(N)}_{[\alpha]}
    \label{eq:balance}
\end{gather}
for some $N^\text{th}$ order tensor field $\mathcal{M}_{[\alpha]}^{(N)}$ with source term $S_{[\alpha]}^{(N)}$, where $[\alpha]$ designates a set of indices $\alpha_1, \ldots ,\alpha_N$.
As shown in \textcite{farag_consistency_2021}, such chains of balance equations can be computed by LBMs, which mimic an implicit Crank-Nicolson scheme with second order accuracy, where $\mathcal{M}_{[\alpha]}^{(N)}$ is identified with the $N^\text{th}$ order moment of the distribution function.

Consequently, interpreting the temporal and spatial derivatives in Eq.~\eqref{eq:balance} as being defined in the reference (Lagrangian) configuration, the LBM can be used to solve conservation laws.
The first three terms in such a moment chain can be written as 
\begin{align}
    & \mathrm{d}_t \mathcal{M}^{(0)} + \mathrm{d}_\alpha \mathcal{M}^{(1)}_{\alpha} = S^{(0)},\nonumber\\
    & \mathrm{d}_t \mathcal{M}^{(1)}_{\alpha} + \mathrm{d}_\beta \mathcal{M}^{(2)}_{\alpha\beta} = S^{(1)}_{\alpha},\nonumber\\
    & \mathrm{d}_t \mathcal{M}^{(2)}_{\alpha \beta} + \mathrm{d}_\gamma \mathcal{M}^{(3)}_{\alpha\beta\gamma} = S^{(2)}_{\alpha\beta},
    \label{eq:chain}
\end{align}
where the moments are computed from
\begin{align}
    & \mathcal{M}^{(0)} = \sum_q {f}_q + \frac{1}{2} S^{(0)} \Delta t \nonumber,\\
    & \mathcal{M}^{(1)}_{\alpha} = \sum_q c_{q\alpha} {f}_q + \frac{1}{2} S^{(1)}_{\alpha} \Delta t \nonumber , \\
    & \mathcal{M}^{(2)}_{\alpha \beta} = \sum_q c_{q\alpha} c_{q\beta} {f}_q + \frac{1}{2} S^{(2)}_{\alpha\beta} \Delta t.
    \label{eq:moment}
\end{align}
These moments give the macroscopic observables and are used in the definition of the equilibrium distribution $f^\text{eq}(\mathcal{M}^{(0)}, \vec{\mathcal{M}}^{(1)}, \vec{\mathcal{M}}^{(2)}, \dots)$.
Likewise, the source terms can be regarded as contributions to a broader forcing term $\psi(S^{(0)}, \vec{S}^{(1)}, \vec{S}^{(2)}, \dots)$.

Eq.~\eqref{eq:chain} describes balance laws, which can also be interpreted with respect to material rather than spatial coordinates.
This is more convenient for solid simulations, especially at large deformations, since in general solids do not assume the shape of the vessel containing them.
Thus, boundary conditions must be formulated with respect to parts of the material rather than regions in space.
Furthermore, material time derivatives appearing in solid-mechanical balance laws yield convective terms in the Eulerian framework, but not in the Lagrangian setting \cite{holzapfel_non-linear_2000}.
This fact simplifies derivations considerably.

\section{Lattice Boltzmann for solid mechanics}
\label{sec:solidlbm}
So far, the treatment of LBMs for balance laws was not bound to any type of material. 
This section deals specifically with solids and introduces the moment chain for non-linear solid mechanics in the reference configuration. 
The final result is the numerical LB scheme.
Boundary conditions are also discussed.

\subsection{A moment-chain for solid mechanics in the reference configuration}
\label{sec:solidlbm:momchain}
Unless it is possible to express the rate of the stress tensor as the divergence of a higher order tensor -- which to the knowledge of the authors is not possible -- the moment $\bm{\mathcal{M}}^{(2)}$ cannot represent the full stress state of the solid.
The remaining part of the constitutive law has to be split off and incorporated into the LBM via a source term.

Previously, for linear elastic materials~\cite{escande_lattice_2020}, the mass density $\rho$, the linear momentum density $\vec{j}$ and the momentum-flux tensor $\vec{\Pi}$ have been chosen as the moments of the LB scheme, with a source term $\vec{S}$ derived from hypoelastic considerations~\cite{murthy_lattice_2018}.
The solid is characterised by three parameters, which will be the reference density $\rho_0$ and the Lamé parameters $\lambda$ and $\mu$.

Here, the moments need to be chosen to form a chain of balance equations, which should resemble the basic equations of solid mechanics: the conservation of mass, the balance of linear momentum and the material law.
The first and second order moments are again chosen as $\vec{j}$ and $\vec{\Pi}$, while the stress tensor is chosen to be the first Piola-Kirchhoff-stress~$\vec{P}$.
This stress measure relates to the reference configuration and enables formulation of the moment chain with physically relevant quantities.
The balance of linear momentum is then given by
\begin{gather}
    \mathrm{d}_t j_\alpha + \mathrm{d}_\beta P_{\alpha \beta} = \rho_0 b_\alpha,
\end{gather}
with external body forces $b_\alpha$.
The LBM requires a source term for the constitutive relation, i.e. for the relation of $\vec{\Pi}$ to $\vec{P}$.
Additionally,
\begin{gather}
    r = \rho_0 \mathrm{d}_\alpha u_\alpha
\end{gather}
is chosen as the moment of $0^\text{th}$ order, representing information on tensile and compressive deformation by the divergence of the displacement field $\vec{u}$.
This is similar to the choice of $\rho$ in the linear elastic method~\cite{faust_dirichlet_2024}. No sources of mass or stress will be considered here, i.e.\ $S^{(0)} \equiv 0$ and $\vec{S}^{(2)} \equiv \vec{0}$.

A $3^\text{rd}$ order moment can be defined~\cite{murthy_lattice_2018} as
\begin{gather}\label{eq:Qlin}
    {Q}_{\alpha \beta \gamma} = c_s^2 ( {j}_\alpha \delta_{\beta \gamma} + {j}_\beta \delta_{\alpha \gamma} + {j}_\gamma \delta_{\alpha \beta} ),
\end{gather}
where $c_s = \sqrt{\sfrac{\mu}{\rho_0}}$ is the material's shear wave velocity.

A material law is required to account for the evolution of the first Piola-Kirchhoff stress tensor $\vec{P}$.
Here, elastic behaviour of the form $\vec{P}(\vec{H})$ is assumed, with the displacement gradient given by
\begin{gather}\label{eq:defgrad}
    H_{\alpha \beta} = \mathrm{d}_\beta u_\alpha.
\end{gather}
The constitutive source term has to capture any part of the stress tensor, that is not covered by the moment $\vec{\Pi}$.
To this end, the difference between
    $P_{\alpha \beta}(\vec{H}) - (- \vec{\Pi}_{\alpha \beta}(\vec{H}))$
is evaluated.
Note that both definitions of stresses differ in signs by convention.

This leads to the source term 
\begin{gather}
    \vec{S}^{(1)} = \rho_0 \vec{b} + \mathrm{div}(\vec{P}(\vec{H}) + \vec{\Pi} (\vec{H})).
    \label{eq:source}
\end{gather}
Since no other source terms are considered in this method, the notation will be simplified from here on, i.e.\ ${\vec{S}^{(1)} \equiv \vec{S}}$.

This approach is similar to the formulation of stress in particle-based systems~\cite{murdoch_physical_2012}, which is closely linked to the origins of the LBM in kinetic theory.
The stress tensor comprises two contributions:
the first is of thermo-kinetic nature and represents the momentum flux.
The second term models particle interactions, where its divergence directly gives rise to interaction forces, represented by \eqref{eq:source}.

With the above, a moment chain formulation for non-linear solid mechanics in the reference configuration has been identified.
This is given by
\begin{align}
    & \mathrm{d}_t r + \mathrm{d}_\alpha {j}_\alpha = 0, \nonumber \\
    & \mathrm{d}_t {j}_\alpha + \mathrm{d}_\beta \Pi_{\alpha \beta} = \rho_0 b_\alpha + \mathrm{d}_\beta \left( P_{\alpha \beta}(\vec{H})+ \Pi_{\alpha \beta}(\vec{H}) \right), \nonumber \\
    & \mathrm{d}_t \Pi_{\alpha \beta} + \mathrm{d}_\gamma {Q}^\text{eq}_{\alpha \beta \gamma} = 0.
    \label{eq:chainmod}
\end{align}
The $3^\text{rd}$ order tensor $\vec{Q}$ has to be taken at equilibrium in order to find closure for the chain of equations~\cite{succi_lattice_2018}.

Note that $\vec{S}$ must be evaluated at time $t$ in the collision step, but at time $t+\Delta t$ in the moment computation, for some iteration starting at time $t$.
Since the displacement at time $t+\Delta t$ is unknown prior to the moment computation, $\vec{S}$ is instead evaluated at time $t$ in Eq.~(\ref{eq:moment}). Thus, this scheme does not mimic a fully explicit second-order Crank-Nicolson scheme for the non-linear component of the material law, but it does for the part contained in $\vec{\Pi}$).
The approximation error resulting from this operation is of order $\mathcal{O}(\Delta t^2)$.
When using the trapezoidal rule to obtain $\vec{u}$ at $t+\Delta t$ from $\vec{j}$, the error reduces to $\mathcal{O}(\Delta t^3)$.

\subsection{The LBM for non-linear solids}
\label{sec:solidlbm:nonlin}
For the moment chain of a solid body in the reference configuration, the moments $\mathcal{M}^{(0)}=r$, $\vec{\mathcal{M}}^{(1)}=\vec{j}$, and $\vec{\mathcal{M}}^{(2)}=\vec{\Pi}$ were selected.
The equilibrium distribution function~\cite{murthy_lattice_2018}
\begin{gather}
    f_q^{eq} = w_q \left( r + \frac{1}{c_s^2} c_{q \alpha} j_\alpha + \frac{1}{2 c_s^4} ( \Pi_{\alpha \beta} - r \, c_s^2 \delta_{\alpha \beta} ) ( c_{q \alpha} c_{q \beta} - c_s^2 \delta_{\alpha \beta} ) \right)
\end{gather}
yields precisely the tensors $r$, $\vec{j}$, and $\vec{\Pi}$ via the moments,
\begin{align}
    r&=\sum_q f_q^\text{eq}, \nonumber\\
    {j}_\alpha & = \sum_q f_q^\text{eq} c_{q \alpha}, \nonumber\\
    \Pi_{\alpha \beta} & = \sum_q f_q^\text{eq} c_{q \alpha} c_{q \beta}, 
    \label{eq:moments}
\end{align}
which guarantees that the moments of ${f}_q$ behave as in Eq.~\eqref{eq:moment}~\cite{farag_consistency_2021}.
The weights for the D2Q9-scheme are given in Tab.~\ref{tab:balance:scheme}.
The forcing contribution~\cite{murthy_lattice_2018},
\begin{gather}\label{eq:forcing}
    \psi_q = w_q \frac{1}{c_s^2} c_{q \alpha} S_\alpha,
\end{gather}
with the source term from Eq.~\eqref{eq:source}, accounts for the remainder of the material law.
Because the displacement vector $\vec{u}$ is not directly available in the LBM algorithm, it must be computed from the linear momentum density by integration. Here, a trapezoidal rule
\begin{gather}
    u_\alpha(\vec{x},t+\Delta t) = u_\alpha(\vec{x},t) + \frac{1}{2\rho_0} \left( {j}_\alpha(\vec{x},t+\Delta t)+{j}_\alpha(\vec{x},t) \right)
\end{gather}
is used.
The displacement gradient $H_{\alpha \beta}=\mathrm{d}_\beta u_\alpha$ is than determined from the displacements and is used to evaluate the first Piola-Kirchhoff stress tensor $\vec{P}$ in the source term \eqref{eq:source}.
For the gradient and divergence computations, second-order central finite difference (FD) stencils are utilized in the interior and first-order stencils at boundary nodes.

Finally, it should be noted that the lattice velocity $c$, which determines the discretisation, is related to the shear velocity by $c = \sqrt{3} \, c_s$.
This prefactor arises from theoretical considerations regarding the anisotropy of the discrete lattice~\cite{escande_lattice_2020}.
The speed of information on the lattice is directly related to $c$.
Regarding the Courant-Friedrichs-Lewy (CFL) condition, there are two wave speeds to be considered in a solid:
the shear wave speed  $c_s$ and the dilatational wave speed $c_d = \sqrt{\sfrac{(\lambda + 2 \mu)}{\rho_0}}$, where $c_s < c_d$.
For stability, the CFL condition requires that the maximum physical information speed, $c_d$, must not exceed the maximum information speed of the numerical scheme, $c_\text{max}$.
In \textcite{escande_lattice_2020}, it is argued that $c_\text{max} = 2c$, due to the second-order FD stencils.
Here, however, only first-order stencils are used at boundary nodes, resulting in $c_d \leqslant c$ as the CFL condition.
This condition can be expressed as
\begin{gather}
    c_d \leqslant 3 c_s,
    \label{eq:CFL}
\end{gather}
which imposes a constraint on the material parameters.
This inequality is satisfied for Poisson's ratio $\nu \leqslant 0.25$.
With second-order FD, $cd \leqslant 2 c = 6 c_s$ and thus $\nu \leqslant \sfrac{5}{11}$.
This upper bound is closer to the incompressible limit, $\nu = 0.5$, in solids.

For the relaxation time, $\tau > 0.5$ is needed for stability.
In the case of the BGK collision operator~\eqref{eq:balance:bgk}, $\tau = 0.55$ is suggested by \textcite{escande_lattice_2020} for the linear elastic scheme, which is adopted in \textcite{faust_dirichlet_2024} as well.

The complete LB algorithm used to solve the solid mechanical moment chain in (\ref{eq:chainmod}) is reported in algorithm \ref{alg:solidlbm}. In the next section, a brief discussion of handling boundary conditions is presented.  

\begin{algorithm}
    \caption{The proposed LB algorithm for non-linear solid mechanics.}
    \label{alg:solidlbm}
    
    \DontPrintSemicolon
    
    \SetKwBlock{Set}{Input}{}
    \SetKwBlock{Init}{Initialise}{}

    \Set{
        lattice nodes $X \in L$\;
        boundaries $\partial L_u$, $\partial L_t$\;
        material parameters $\lambda$, $\mu$, $\rho_0$\;
        numerical parameter $\tau$\;
    }
    \Init{
        material parameter $c_s$\;
        lattice parameters $c_q$, $w_q$, $\Delta X$, $\Delta t$\;
        moments $r$, $\vec{j}$, $\vec{\Pi}$\;
        distributions $f_q \gets f_q^\text{eq}$\;
        fields $\vec{u}$, $\vec{H}$, $\vec{P}$\;
        source $\vec{S}$\;
        time $t \gets 0$\;
    }
    
    \tcp{main simulation loop}
    \While{$t \leq t_{max}$}{
        \ForAll{$X \in L$}{
            \ForEach{$q$}{
                \tcp{update moments and calculate equilibrium distributions}
                $r \gets \sum_q f_q $\;
                $j_\alpha \gets \sum_q c_{q \alpha} f_q + \frac{\Delta t}{2} S_\alpha$\;
                $\Pi_{\alpha \beta} \gets \sum_q c_{q \alpha} c_{q \beta} f_q $\;
                $f_q^\text{eq} \gets w_q \left( r + \frac{1}{c_s^2} c_{q \alpha} {j}_\alpha + \frac{1}{2c_s^4} ( \Pi_{\alpha \beta} - r c_s^2 \delta_{\alpha \beta} ) ( c_{q \alpha} c_{q \beta} - c_s^2 \delta_{\alpha \beta} ) \right)$\;
                \tcp{compute source}
                $\psi_q \gets w_q \frac{1}{c_s^2} c_{q \alpha} S_\alpha$\;
                \tcp{collide}
                ${f}_q^\text{coll} \gets {f}_q - \frac{\Delta t}{\tau} \left( f_q - f_q^\text{eq} \right) + \left( 1 - \frac{\Delta t}{2 \tau} \right) \psi_q \Delta t$\;
                \tcp{stream to neighbours}
                $f_q(X+c_q \Delta t, t+\Delta t) \gets f_q^\text{coll}(X,t)$\;
            }
        }
        \tcp{handle boundaries}
        \ForAll{$X \in \partial L_u$}{
            \ForEach{$q$ crossing $\partial L_u$}{
                $f_{\Bar{q}}(X,t+\Delta t) \gets f_q^\text{coll}(X,t) - w_q \frac{2}{c_s^2} c_{q \alpha} {j}_\alpha^\text{bd}$\;
            }
        }
        \ForAll{$X \in \partial L_t$}{
            \ForEach{$q$ crossing $\partial \Omega_t$}{
                $f_{\Bar{q}}(X,t+\Delta t) \gets -{f}_q^\text{coll}(X,t) + 2 w_q \left( r^\text{bd} + \frac{1}{2 c_s^4}( \Pi_{\alpha \beta}^\text{bd} - r c_s^2 \delta_{\alpha \beta} ) ( c_{q \alpha}c_{q \beta} - c_s^2 \delta_{\alpha \beta} ) \right)$\;
            }
        }
        \tcp{calculate stresses from gradient}
        \ForAll{$X \in L$}{
            $H_{\alpha \beta} \gets \mathrm{d}_\beta u_\alpha$\;
            $P_{\alpha \beta} \gets P_{\alpha \beta}(\vec{H})$\;
            $\Pi_{\alpha \beta} \gets \Pi_{\alpha \beta}(\vec{H})$\;
        }
        \tcp{update displacement and moments}
        \ForAll{$X \in L$}{
            $S_\alpha \gets \rho_0 b_\alpha + \mathrm{d}_\beta ( \Pi_{\alpha \beta} + P_{\alpha \beta} )$\;
            $r \gets \sum_q f_q $\;
            $j_\alpha \gets \sum_q c_{q \alpha} f_q + \frac{\Delta t}{2} S_\alpha$\;
            $u_\alpha(X, t+\Delta t) \gets u_\alpha(X, t) + \frac{1}{2\rho_0} \big( j_\alpha(X, t+\Delta t) + j_\alpha(X,t) \big)$\;
        }
        $t \gets t + \Delta t$\;
    }
\end{algorithm}

\subsection{Boundary conditions}
\label{sec:solidlbm:bc}

In a previous publication~\cite{faust_dirichlet_2024} for linear elastic materials, an adaption of bounce-back and anti-bounce-back boundary rules~\cite{kruger_lattice_2017}) was proposed, in order to approximate the behaviour of Dirichlet and Neumann boundaries within an LBM.
The idea of bounce-back boundary conditions is to send any outgoing distribution $f_q$ back in the opposite direction $\Bar{q}$,
\begin{gather}
    f_{\Bar{q}}(\vec{x},t+\Delta t) = (-) f_q^{\text{coll}}(\vec{x},t),
\end{gather}
where the minus is applied for anti-bounce-back.

These bounce-back rules can be modified by adding a term that depends on the prescribed moments in order to account for non-zero Dirichlet or Neumann boundary conditions.
For Dirichlet, the momentum $\vec{j}^*$ is prescribed.
For Neumann, the prescribed traction $\vec{T}^*$ is used to set the momentum flux $\vec{\Pi}^*$.

Here, an analogous methodology is used to impose boundary values in the reference configuration.
This is accomplished in the linear momentum density $\vec{j}$ via
\begin{gather}\label{eq:bbcon}
    {f}_{\Bar{q}}(\vec{x},t+\Delta t) = {f}_q^{\text{coll}}(\vec{x},t) - \frac{2}{c_s^2} w_q c_{q \alpha} {j}_\alpha^{*},
\end{gather}
for Dirichlet boundary conditions and on the momentum flux tensor $\vec{\Pi}$ via
\begin{gather}\label{eq:abbcon}
    {f}_{\Bar{q}}(\vec{x},t+\Delta t) = -{f}_q^{\text{coll}}(\vec{x},t) + 2 w_q \left( r^{\text{bd}} + \frac{1}{2 c_s^4}( \Pi_{\alpha \beta}^{*} - r^{\text{bd}} c_s^2 \delta_{\alpha \beta} ) ( c_{q \alpha} c_{q \beta} - c_s^2 \delta_{\alpha \beta} ) \right)
\end{gather}
for Neumann boundary conditions.
Boundary values in the Poisson tensor may be obtained from the traction vectors~$\vec{T^*}$ through
\begin{gather}
     P_{\alpha \beta} N_\beta = T_\alpha^*, \nonumber \\
    \Leftrightarrow\; -\Pi_{\alpha \beta} N_\beta + ( P_{\alpha \beta} + \Pi_{\alpha \beta} ) N_\beta = T_\alpha^*, \nonumber \\
    \Rightarrow\; \Pi_{\alpha \beta} N_\beta = - T_\alpha^* + ( P_{\alpha \beta} + \Pi_{\alpha \beta} ) N_\beta.
\end{gather}
As in \cite{faust_dirichlet_2024}, these equations are transformed into a coordinate system normal to the boundary, in which the quasi-boundary tractions $-\vec{T^*}+(\vec{P}+\vec{\Pi})\vec{N}$ may simply be substituted into the first row and column of $\vec{\Pi}$, while the remaining entries are extrapolated to the boundary. The momentum flux and first Piola-Kirchhoff tensors appearing in the second term are also extrapolated to the boundary.

In the same manner as the boundary conditions for linear elasticity, the schemes proposed here for non-linear material models are expected to exhibit first-order accuracy.
This is consistent with the first-order finite difference stencils used at boundary nodes.

\section{Numerical experiments}
\label{sec:num}

To investigate the performance of the proposed scheme and to showcase its capabilities, numerical studies are undertaken.
Material models of type St.\ Venant-Kirchhoff (SVK) and neo-Hooke (NH) are considered here.
The stress-strain relations are
\begin{align}
    S_{ij} &= \lambda E_{kk} \delta_{ij} + 2 \mu E_{ij}
        &\text{(St.\ Venant-Kirchhoff)},
    \\
    S_{ij} &= \frac{\lambda}{2} \left( J^2 -1\right) \, C_{ij}^{-1} + \mu \left( \delta_{ij} - C_{ij}^{-1} \right)
        &\text{(neo-Hooke)},
\end{align}
for the second Piola-Kirchhoff stress tensor $\vec{S}$, where 
$\vec{E} = \frac{1}{2} \left( \vec{H}^T + \vec{H} + \vec{H}^T \vec{H} \right)$,
$\vec{C} = (\vec{H} + \vec{1})^T (\vec{H} + \vec{1})$, and
$J = \det (\vec{H} + \vec{1})$,
in terms of the displacement gradient $\vec{H}$.
The first Piola-Kirchhoff stress tensor $\vec{P}$, as used in Sect.~\ref{sec:solidlbm}, is computed by $\vec{P} = (\vec{H} + \vec{1}) \, \vec{S}$.

At first, two benchmark models are investigated: uniaxial tension and simple shear.
The boundary conditions are applied at a slow rate, ensuring the system primarily exhibits quasi-static characteristics with only a weakly dynamic effect due to inertia forces.
Next, the propagation of bending waves in a beam is simulated to show the performance in a dynamical system.
For all simulations, Finite Element (FE) results are obtained for comparison.
Theses FE simulations are carried out with the open source project \textit{FEniCSx}~\cite{baratta_dolfinx_2023}.

From the LB data and the FE reference, deviations $\delta \vec{u} = \vec{u}_\text{LB} - \vec{u}_\text{FE}$ are evaluated at each lattice node.
From this, relative grid errors
\begin{align}
    &\mathcal{E}_2 = \frac{1}{n} \frac{\Vert \delta \vec{u} \Vert_2}{\Vert \vec{u} \Vert_2}, &\;
    &\mathcal{E}_\infty = \frac{\Vert \delta \vec{u} \Vert_\infty}{\Vert \vec{u} \Vert_2}, &\;
    \label{eq:num:griderror}
    \intertext{where}
    &\Vert \vec{u} \Vert_2 \equiv \sqrt{\sum_\text{lattice} \boldsymbol{u} \cdot \boldsymbol{u}}, &\;
    &\Vert \vec{u} \Vert_\infty \equiv \max_\text{lattice} (u_1, u_2),
    \label{eq:num:gridnorm}
\end{align}
are computed.
Herein, $n$ is the number of nodes.
Eq.~\eqref{eq:num:gridnorm} describes the grid norm over all lattice nodes for either the displacement $\vec{u}$ or its deviation $\delta \vec{u}$ with reference to the 2-norm and the maximum norm.
$\mathcal{E}_2$ represents the relative mean error across the full grid.
The resulting error measures in Eq.~\eqref{eq:num:griderror} are considered for the validation of the algorithm.
For simplicity, any parameters are considered to be dimensionless in the following.

\subsection{Uniaxial tension}
\label{sec:num:tension}

\begin{figure}[tp]
    \centering
    \subcaptionbox{domain\label{fig:num:tension:setup:domain}}
        {\includegraphics[width=4cm]{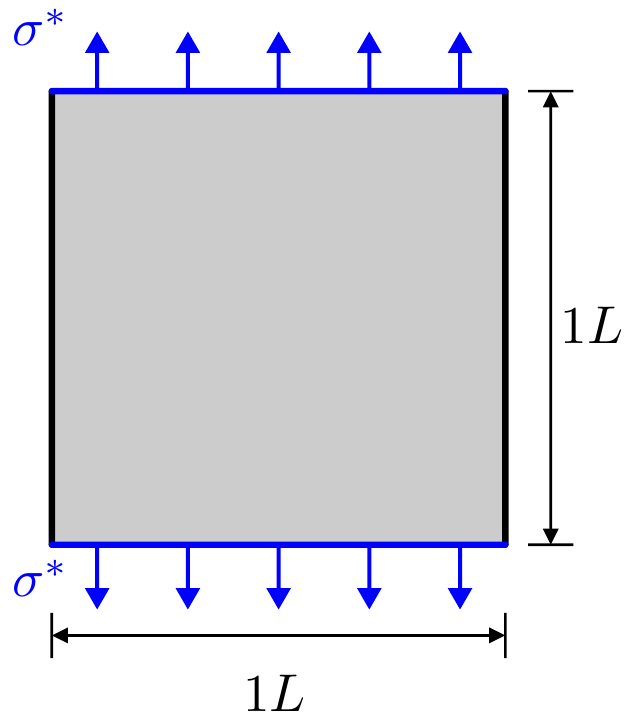}}
    \subcaptionbox{boundary condition\label{fig:num:tension:setup:load}}
        {\input{neumann_bc.pgf}}
    \caption{
        The (a) domain and boundary conditions and (b) the value of the applied traction for the uniaxial tension benchmark.
    }
    \label{fig:num:tension:setup}
\end{figure}

The first benchmark problem is a uniaxial tension test.
Traction is applied at the top and bottom of a square domain by a Neumann boundary condition, see Fig.~\ref{fig:num:tension:setup:domain}.
The left and right boundaries are free with a homogeneous Neumann boundary condition.
The domain has an extension of $1L$ by $1L$, where $L$ is an arbitrary unit of length.
A lattice spacing of $\Delta x = 0.025 L$ is chosen, resulting in a total number of $1\,600$ lattice nodes.
For a convergence study, additional lattice spacings $\Delta x \in \{ 0.05 L, 0.0125 L, 0.006125 L \}$ are introduced.
The traction is slowly increased and then kept constant, see Fig.~\ref{fig:num:tension:setup:load}, by the time-dependent function
\begin{align}
    \sigma^\ast = \alpha \, \begin{cases}
        \sin^2 \left( \tfrac{\pi}{4} t \right), &t < 2, \\
        1,                                      &t \geqslant 2.
    \end{cases}
    \label{eq:num:tension:bc}
\end{align}
This ensures a continuous and almost quasi-static transition from zero to a constant value.
The parameter $\alpha$ scales the amplitude depending on the material law, where
\begin{align}
    \alpha &= 0.175 \quad \text{for St.\ Venant-Kirchhoff and} \notag\\
    \alpha &= 0.35 \quad \text{for neo-Hooke}. 
\end{align}

A range of values for the Poisson ratio $\nu$ is considered with $\nu \in \{0, 0.1, 0.2, 0.25, 0.3\}$.
In the case of the St. Venant-Kirchhoff material, the deformations are small enough to allow for a comparison to the linear elastic model presented in \textcite{faust_dirichlet_2024}, where $\nu = 0.25$ is chosen due to stability considerations.
For the neo-Hooke material, auxetic material behaviour with $\nu = -0.1$ is additionally considered.
The mass density is set to $\rho_0 = 1$.
Lastly, the relaxation time $\tau$, cf. the LBE~\eqref{eq:balabce:lbe-coll}, is chosen as $\tau = 0.55$.

\begin{figure}
    \centering
    \subcaptionbox{St.\ Venant-Kirchhoff, $\nu = 0.25$\label{fig:num:tension:deform:svk}}
        {\includegraphics[width=0.45\textwidth]{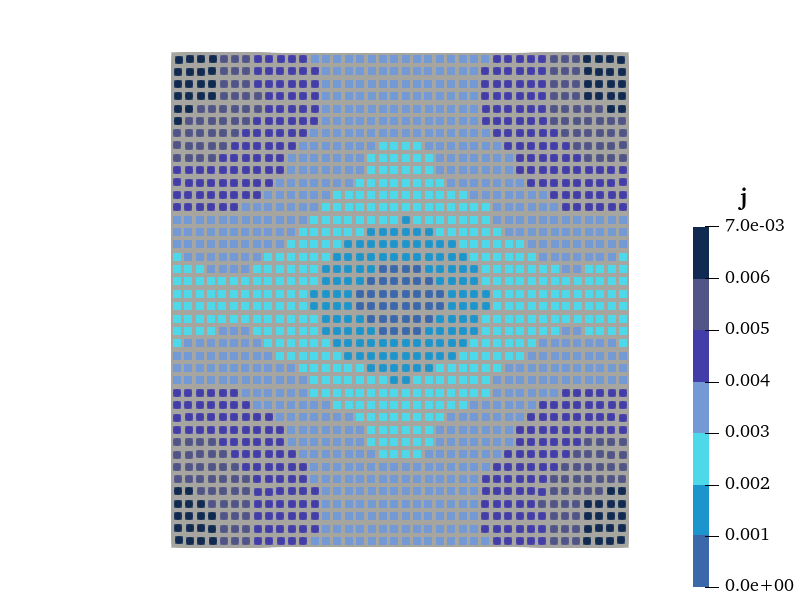}}
    \subcaptionbox{neo-Hooke, $\nu = -0.10$\label{fig:num:tension:deform:nh}}
        {\includegraphics[width=0.45\textwidth]{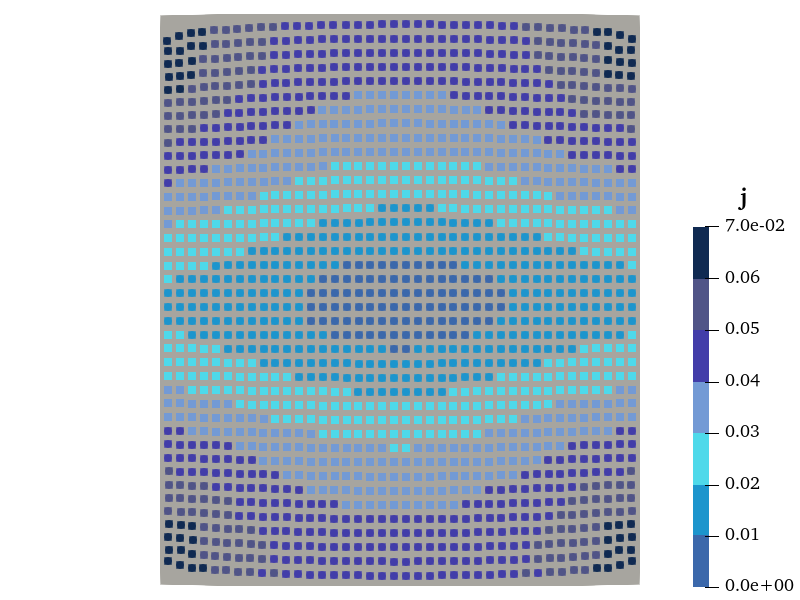}}
    \caption{
        Two examples of deformed domains for the uniaxaial tension benchmark with
        (a) St.\ Venant-Kirchhoff and (b) neo-Hooke material behaviour at $t = 2.2$.
        In the background is the deformed domain of the FE solution overlaid with points representing the LB nodes.
        The points are coloured according to the magnitude of the momentum $\vec{j}$.
    }
    \label{fig:num:tension:deform}
\end{figure}

\begin{table}[tbp]
    \begin{minipage}{0.4\textwidth}
        \caption{The grid errors for the uniaxial tension benchmark for Saint Venant-Kirchhoff (SVK), Neo-Hooke (NH) and linear elastic (LE) material behaviour.}
        \label{tab:num:tension:errors}
        \begin{tabular}{@{}lrrr@{}}
            \toprule
            \rowcolor[HTML]{B2B2B2} 
            model & $\nu$    & $\mathcal{E}_2$      & $\mathcal{E}_\infty$    \\
            \midrule
            SVK   & 0.30  & 7.6E-05 & 6.9E-03 \\
            \rowcolor[HTML]{EEEEEE} 
            SVK   & 0.25  & 3.7E-05 & 5.4E-03 \\
            SVK   & 0.20  & 3.7E-05 & 5.7E-03 \\
            \rowcolor[HTML]{EEEEEE} 
            SVK   & 0.10  & 4.3E-05 & 6.5E-03 \\
            SVK   & 0.00  & 7.8E-05 & 8.7E-03 \\
            \rowcolor[HTML]{EEEEEE} 
            LE    & 0.25  & 2.9E-05 & 5.4E-03 \\
            \midrule
            NH    & 0.30  & 4.4E-05 & 6.7E-03 \\
            \rowcolor[HTML]{EEEEEE} 
            NH    & 0.25  & 5.0E-05 & 7.5E-03 \\
            NH    & 0.20  & 5.3E-05 & 8.2E-03 \\
            \rowcolor[HTML]{EEEEEE} 
            NH    & 0.10  & 6.6E-05 & 1.0E-02 \\
            NH    & 0.00  & 8.6E-05 & 1.3E-02 \\
            \rowcolor[HTML]{EEEEEE} 
            NH    & -0.10 & 1.1E-04 & 1.8E-02 \\
            \bottomrule
        \end{tabular}
    \end{minipage}
    \hfil
    \begin{minipage}{0.4\textwidth}
        \caption{The grid errors for the simple shear benchmark for Saint Venant-Kirchhoff (SVK), Neo-Hooke (NH) and linear elastic (LE) material behaviour.}
        \label{tab:num:shear:errors}
        \begin{tabular}{@{}lrrr@{}}
            \toprule
            \rowcolor[HTML]{B2B2B2} 
            model & $\nu$    & $\mathcal{E}_2$      & $\mathcal{E}_\infty$    \\
            \midrule
            SVK   & 0.30  & 8.9E-06 & 9.5E-04 \\
            \rowcolor[HTML]{EEEEEE} 
            SVK   & 0.25  & 6.3E-06 & 1.6E-03 \\
            SVK   & 0.20  & 1.1E-05 & 1.1E-03 \\
            \rowcolor[HTML]{EEEEEE} 
            SVK   & 0.10  & 1.0E-05 & 1.7E-03 \\
            SVK   & 0.00  & 1.5E-05 & 1.6E-03 \\
            \rowcolor[HTML]{EEEEEE} 
            LE    & 0.25  & 4.4E-05 & 3.1E-02 \\
            \midrule
            NH    & 0.30  & 5.3E-06 & 8.9E-04 \\
            \rowcolor[HTML]{EEEEEE} 
            NH    & 0.25  & 6.3E-06 & 9.7E-04 \\
            NH    & 0.20  & 7.6E-06 & 1.0E-03 \\
            \rowcolor[HTML]{EEEEEE} 
            NH    & 0.10  & 1.0E-05 & 1.1E-03 \\
            NH    & 0.00  & 1.3E-05 & 1.4E-03 \\
            \rowcolor[HTML]{EEEEEE} 
            NH    & -0.10 & 1.6E-05 & 1.7E-03 \\
            \bottomrule
        \end{tabular}
    \end{minipage}
\end{table}

\begin{figure}
    \centering
    \subcaptionbox{St.\ Venant-Kirchhoff\label{fig:num:tension:svk}}
        {\input{convergence_tension_svk.pgf}}
    \hfil
    \subcaptionbox{neo-Hooke\label{fig:num:tension:nh}}
        {\input{convergence_tension_nh.pgf}}
    \caption{
        The mean ($\mathcal{E}_2$) and maximum ($\mathcal{E}_\infty$) grid error for the lattice refinement studies of the unixaxial tension benchmark.
        (a) St.\ Venant-Kirchhoff and (b) neo-Hooke material, each with $\nu = 0.20$.
        Also shown for comparison are lines indicating linear and quadratic convergence rates.
    }
    \label{fig:num:tension}
\end{figure}

The applied loading deforms and elongates the domain, see Fig.~\ref{fig:num:tension:deform} for examples.
After $t = 2$, the traction is kept constant.
However, the simulation does show some inertial effects, since the momentum in Fig.~\ref{fig:num:tension:deform} at $t = 2.2$ is not at zero throughout the domain, so the displacement field is not stationary.
The domain undergoes oscillations around the stationary state.
The displacement of the LB nodes generally follows the FE results.
Due to the scaling of the traction load, the SVK simulations exhibit small deformations.
At larger deformations the SVK model can become unstable.
In contrast, the domain exhibits larger displacements with the neo-Hooke material without instabilities.
Deviations can be observed along the edges towards the corners, especially for the neo-Hooke material, see Fig.~\ref{fig:num:tension:deform:nh}, where significant differences between LB and FE solution occur.

The grid errors according to Eq.~\eqref{eq:num:griderror} at time $t = 2.2$ are given in Tab.~\ref{tab:num:tension:errors}.
For both St. Venant-Kirchhoff and neo-Hooke material the maximum error is still relatively low with $\mathcal{E}_\infty < 2 \times 10^{-2}$ and the mean error $\mathcal{E}_2$ is about 2 orders of magnitude smaller.
The errors increase with a decrease in $\nu$ and thus the largest errors are found for the neo-Hooke material with $\nu = -0.10$.
For the linear elastic (LE) material law, the error is comparable to the St. Venant-Kirchhoff material with $\nu = 0.25$.

With respect to refinement of the lattice spacing, Fig.~\ref{fig:num:tension} shows the results for both material models.
In both cases the maximum error $\mathcal{E}_\infty$ shows approximately a first-order convergence, whereas the mean error $\mathcal{E}_2$ decreases quadratically.
For the coarsest lattice, $\mathcal{E}_\infty < 2 \times 10^{-2}$ and with $h = 0.025$, which was used in all simulations except the refinement studies, the maximum error is below 1 \%.

\subsection{Simple shear}
\label{sec:num:shear}

\begin{figure}[tp]
    \centering
    \subcaptionbox{domain\label{fig:num:shear:setup:domain}}
        {\includegraphics[width=4cm]{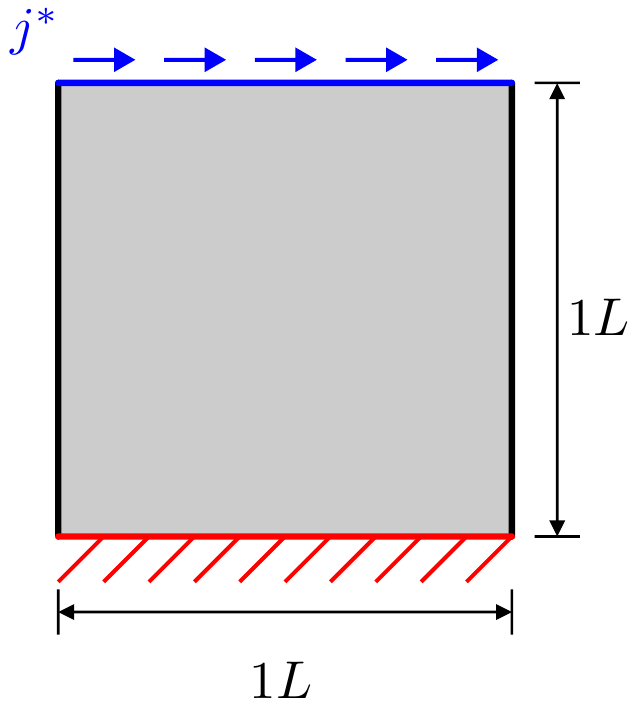}}
    \subcaptionbox{boundary condition\label{fig:num:shear:setup:load}}
        {\input{dirichlet_bc.pgf}}
    \caption{
        The (a) domain and boundary conditions and (b) the value of the prescribed momentum $j^\ast$ for the simple shear benchmark.
        Additionally, the resulting displacement $u^\ast$ is shown as well.
        }
    \label{fig:num:shear:setup}
\end{figure}

For the simple shear benchmark, the dimensions of the domain are chosen as for the tension test.
The boundary at the bottom of the domain is kept fixed via a homogeneous Dirichlet boundary condition.
The top of the domain is shifted to the right by prescribing the momentum by
\begin{align}
    j^\ast = \alpha \, \begin{cases}
        \sin \left( \tfrac{\pi}{2} t \right),   &t < 2, \\
        0,                                      &t \geqslant 2,
    \end{cases}
    \label{eq:num:shear:bc-j}
\end{align}
which results in a displacement
\begin{align}
    u^\ast = \alpha \, \frac{2}{\pi }\, \begin{cases}
        \Big( 1 - \cos \left( \tfrac{\pi}{2} t \right)\Big),    &t < 2, \\
        2,                                                      &t \geqslant 2,
    \end{cases}
    \label{eq:num:shear:bc-u}
\end{align}
which, as for the tension example, transitions continuously and almost quasi-statically from zero to a constant value.
Eq.~\ref{eq:num:shear:bc-u} is used as the boundary condition for the FE simulation.
The amplitude is scaled by
\begin{align}
    \alpha &= 0.03 \quad \text{for St.\ Venant-Kirchhoff and} \notag\\
    \alpha &= 0.1\phantom{0} \quad \text{for neo-Hooke}. 
\end{align}

The remaining parameters for the benchmark simulations and the refinement study, especially $\nu$ and $\rho_0$ for the material and $\Delta x$ and $\tau$ for the numerics, are chosen as in Sect.~\ref{sec:num:tension}.
The linear elastic case with $\nu = 0.25$ is also considered.

\begin{figure}[tp]
    \centering
    \subcaptionbox{St.\ Venant-Kirchhoff, $\nu = 0$\label{fig:num:shear:deform:svk}}
        {\includegraphics[width=0.45\textwidth]{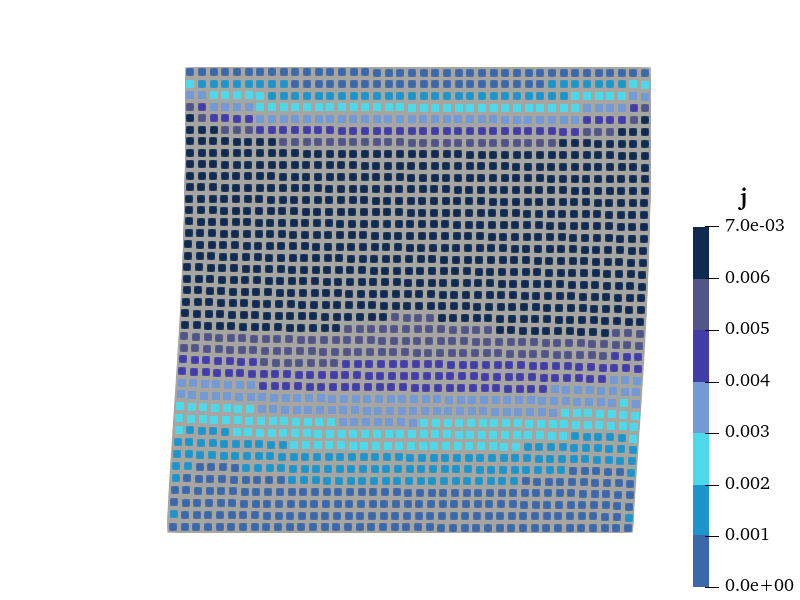}}
    \subcaptionbox{neo-Hooke, $\nu = 0.2$\label{fig:num:shear:deform:nh}}
        {\includegraphics[width=0.45\textwidth]{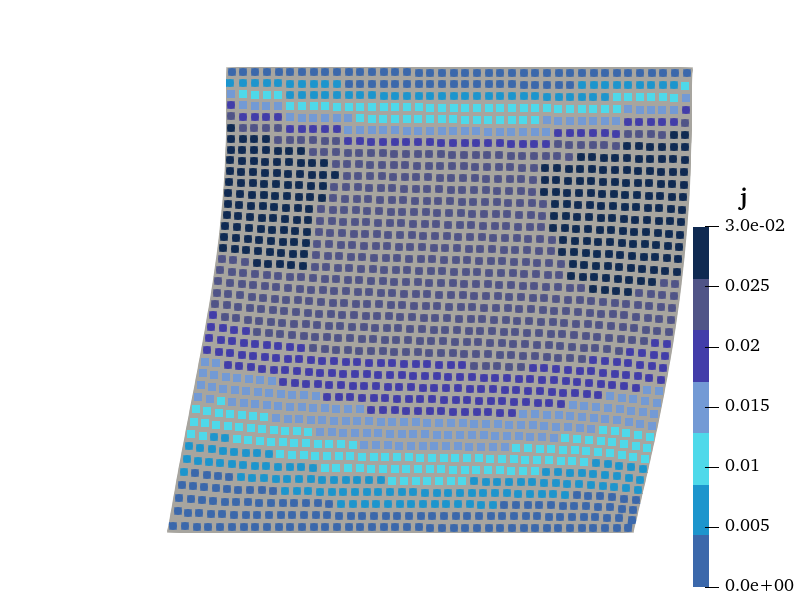}}
    \caption{
        Two examples of deformed domains for the simple benchmark with
        (a) St.\ Venant-Kirchhoff and (b) neo-Hooke material behaviour at $t = 2.2$.
        In the background is the deformed domain of the FE solution overlaid with points representing the LB nodes.
        The points are coloured according to the magnitude of the momentum $\vec{j}$.
    }
    \label{fig:num:shear:deform}
\end{figure}

\begin{figure}[tp]
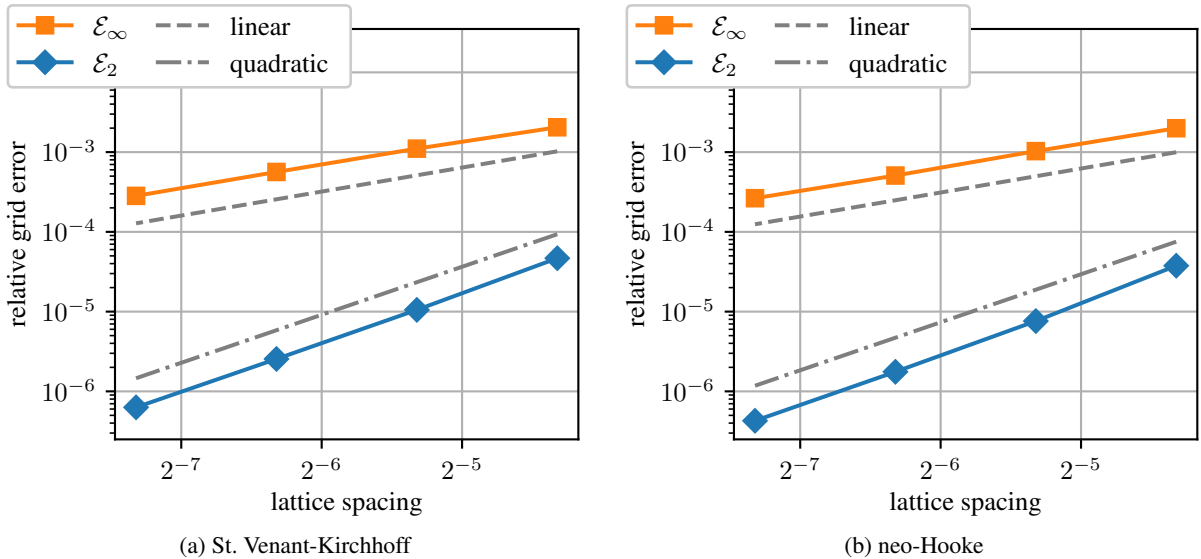

    \centering
    \subcaptionbox{St.\ Venant-Kirchhoff\label{fig:num:shear:svk}}
        {\input{convergence_shear_svk.pgf}}
    \hfil
    \subcaptionbox{neo-Hooke\label{fig:num:shear:nh}}
        {\input{convergence_shear_nh.pgf}}
    \caption{
        The mean ($\mathcal{E}_2$) and maximum ($\mathcal{E}_\infty$) grid error for the lattice refinement studies of the simple shear benchmark.
        (a) St.\ Venant-Kirchhoff and (b) neo-Hooke material, each with $\nu = 0.20$.
        Also shown for comparison are lines indicating linear and quadratic convergence rates.
    }
    \label{fig:num:shear}
\end{figure}

The application of the Dirichlet boundary conditions initially deflects the upper boundary to the right and subsequently fixes the boundary in the deflected position.
Fig.~\ref{fig:num:shear:deform} shows the results for two simulations, St. Venant-Kirchhoff and neo-Hooke material, at $t = 2.2$.
As in the tension example, the deformation does not reach a stationary state.
Instead, oscillations around the equilibrium deformation occur.
The non-zero momentum field, Fig.~\ref{fig:num:shear:deform:nh} clearly shows inertial effects.
The St. Venant-Kirchhoff material only undergoes small deformations, whereas neo-Hooke material shows larger displacements.
Nevertheless, both plots show the same qualitative behaviour.

The displacement of the LB nodes is visually very close to the FE reference in the background.
The error values for all values of $\nu$ are given in Tab.~\ref{tab:num:shear:errors}.
For this example, the errors are smaller compared to the tension example.
$\mathcal{E}_\infty < 1\%$ for St. Venant-Kirchhoff and neo-Hooke material.
$\mathcal{E}_2$ is again at least two orders of magnitude smaller.
The linear elastic (LE) model shows the largest errors for this example, which are one order of magnitude larger than the St. Venant-Kirchhoff with $\nu = 0.25$.

The refinement study for $\nu = 0.20$ again shows a first-order convergence rate for the $\mathcal{E}_\infty$ and a second-order convergence rate for $\mathcal{E}_2$, regardless of the material model.
Here, the maximum error for the coarsest lattice is around~${0.2 \%}$.

\subsection{Bending wave}
\label{sec:num:wave}

\begin{figure}[tp]
    \centering
    \includegraphics[width=12cm]{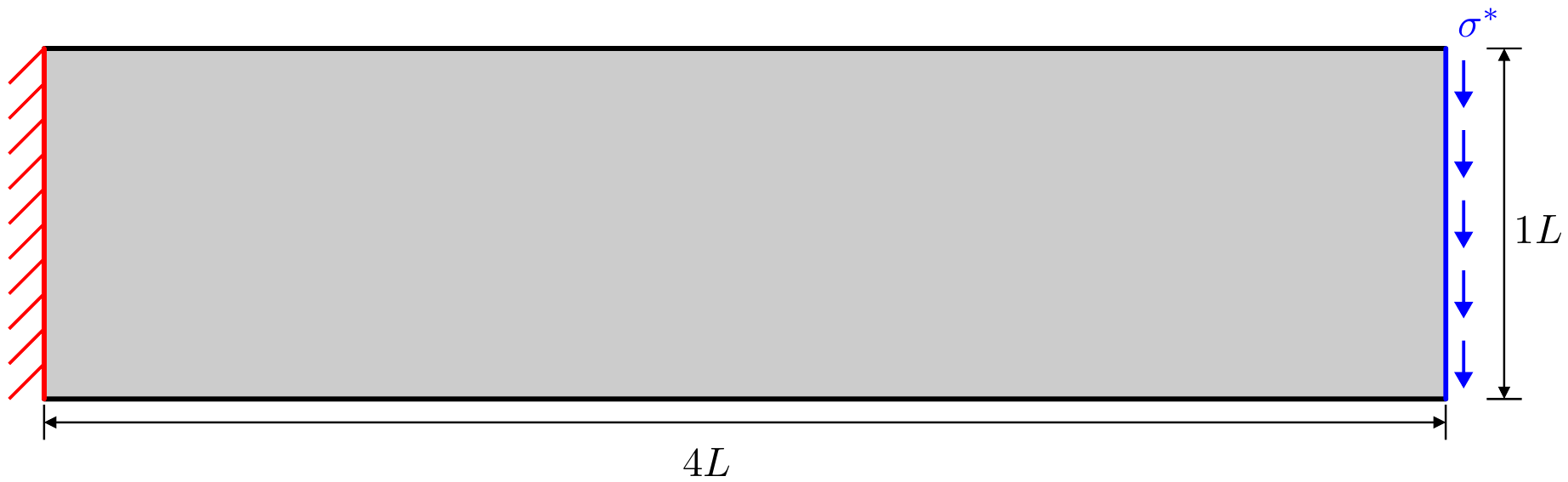}
    \caption{
        The domain for the example of the propagation of a bending wave in a beam.
    }
    \label{fig:num:wave:domain}
\end{figure}

Turning now to a dynamical example, the propagation of bending waves in a cantilever beam is considered.

The domain has extensions $4L$ by $1L$, resembling a short beam, see Fig.~\ref{fig:num:wave:domain}.
It is clamped at the left end, with free boundaries on top and bottom.
A Ricker wavelet~\cite{ricker_form_1953}
\begin{gather}
    \sigma^\ast = \alpha \, \frac{2}{\sqrt{3s \sqrt{\pi}}} \; \left( 1 - \frac{\left(t - t_0\right)^2}{s} \right) \; \exp \left( - \frac{1}{2} \frac{\left(t - t_0\right)^2}{s} \right)
    \label{eq:num:wave:ricker}
\end{gather}
is applied as a traction force via a Neumann boundary condition at the right end of the beam.
The direction is tangential to the boundary, thus exciting shear waves.
The parameters in Eq.~\eqref{eq:num:wave:ricker} are chosen as $\alpha = 0.15$, $s = 0.215$, and $t_0 = 1.0$.
This generates a wave packet with a broad spectrum at the boundary within a time interval $t \in [0, 2]$.
The edge turns into a free boundary for $t>2$.

The neo-Hooke material is chosen with $\nu = 0.20$ and $\rho_0 = 1$.
With regard to the numerics, the lattice spacing is set to $\Delta x = 0.0250$.
The relaxation time is slightly varied, with $\tau \in {0.54, 0.55, 0.57, 0.59}$, in order to investigate the influence on the error.

\begin{figure}[t]
    \def \wavewidth {0.95\textwidth}
    \centering
    \subcaptionbox{$t = 2$\label{fig:num:wave:deform:2}}
        {\includegraphics[width=\wavewidth]{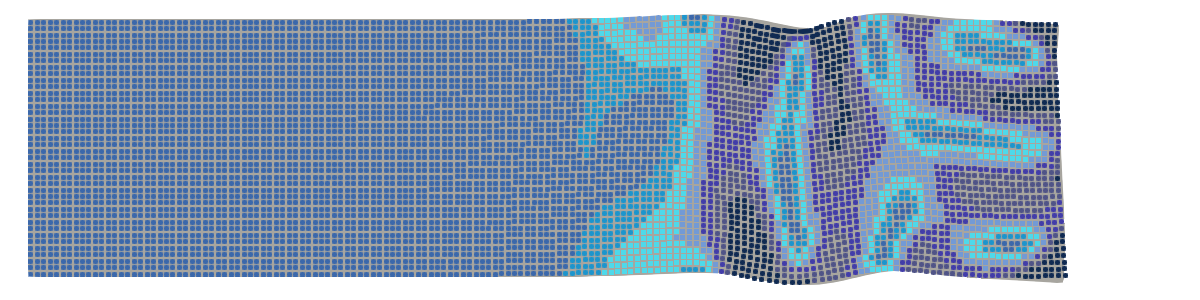}}

    \medskip
    \subcaptionbox{$t = 6$\label{fig:num:wave:deform:6}}
        {\includegraphics[width=\wavewidth]{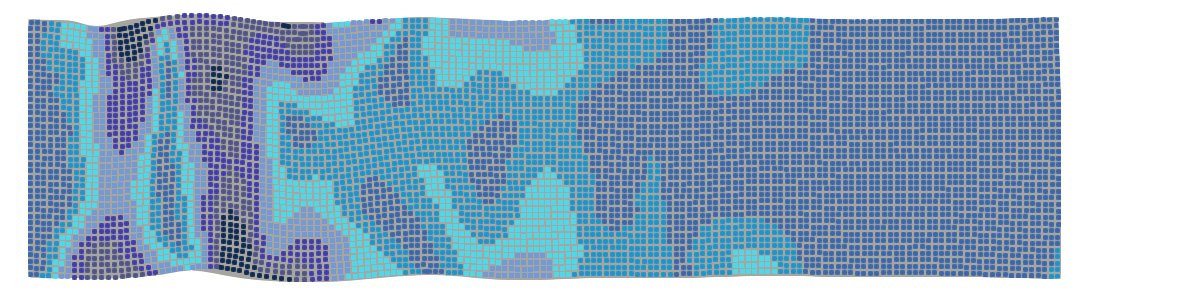}}

    \medskip
    \subcaptionbox{$t = 11$\label{fig:num:wave:deform:11}}
        {\includegraphics[width=\wavewidth]{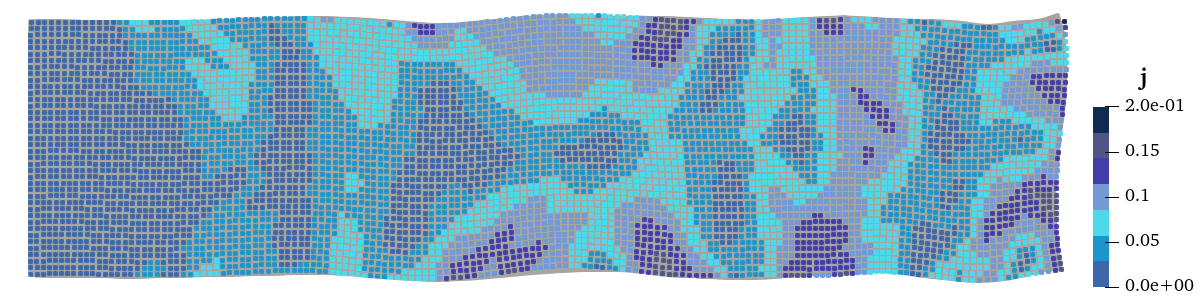}}
    \caption{
        The deformed beam for three different time instances:
        (a) after the application of the Ricker wavelet,
        (b) after reflection at the left end, and
        (c) the dispersed wave packet after reflection at the right end.
        The gray background shows the FE solution with points overlaid, indicating the LB nodes.
        The points are coloured according to the magnitude of the momentum $\vec{j}$.
        The relaxation time is $\tau = 0.55$.
    }
    \label{fig:num:wave:deform}
\end{figure}

\begin{figure}[t]
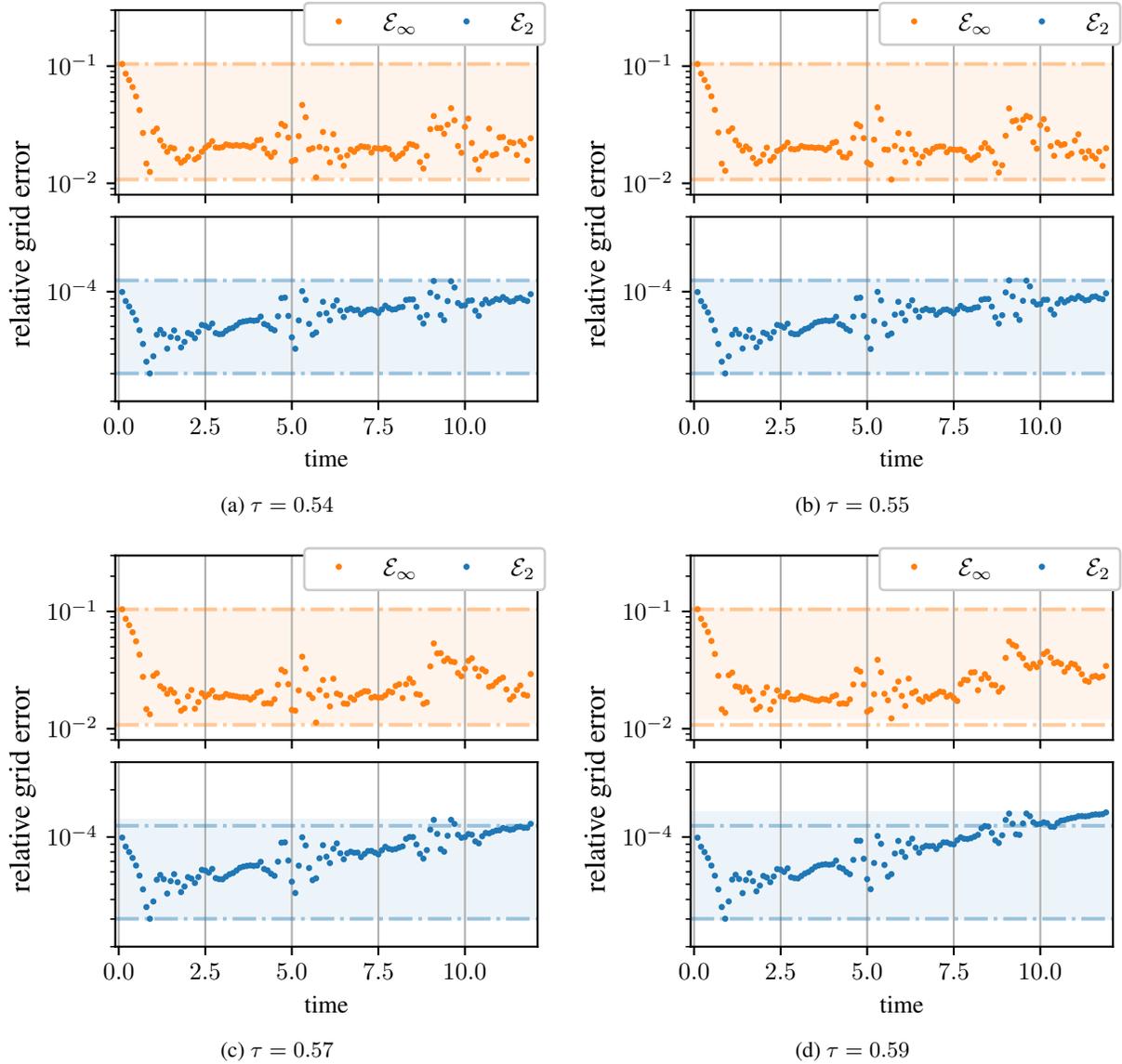

    \centering
    \subcaptionbox{$\tau = 0.54$\label{fig:num:wave:54}}
        {\input{wave_error_tau54.pgf}}
    \hfil
    \subcaptionbox{$\tau = 0.55$\label{fig:num:wave:55}}
        {\input{wave_error_tau55.pgf}}

    \medskip
    \subcaptionbox{$\tau = 0.57$\label{fig:num:wave:57}}
        {\input{wave_error_tau57.pgf}}
    \hfil
    \subcaptionbox{$\tau = 0.59$\label{fig:num:wave:59}}
        {\input{wave_error_tau59.pgf}}
    \caption{
        The grid errors of the wave propagation example for the four different relaxation times (a) $\tau = 0.54$, (a) $\tau = 0.55$, (a) $\tau = 0.57$, and (a) $\tau = 0.59$.
        The horizontal lines indicate the minimum and maximum of the respective data, whereas the shaded area shows the range of errors for $\tau = 0.55$ for comparison.
    }
    \label{fig:num:wave}
\end{figure}

The initial wave packet, in terms of displacement, is comprised of a peak and a trough.
Fig.~\ref{fig:num:wave:deform} shows the deformation for three different time instances with a relaxation time $\tau = 0.55$.
The wave travels to the left, Fig.~\ref{fig:num:wave:deform:2}, is reflected by the clamped end, Fig.~\ref{fig:num:wave:deform:6}, and subsequently reflected by the free end on the right.
At this instance, the wave packet is significantly dispersed, Fig.~\ref{fig:num:wave:deform:11}.

For this example, the errors have been evaluated for a multitude of time instances with $t < 12$ in a longer-running simulation.
The main results are the error plots in Fig.~\ref{fig:num:wave} for the four different relaxation times.
When the wave packet interacts with the boundary, the error values are especially elevated.
In each case the maximum occurs, when the beam is excited.
Further increases are around $t = 5$ and before $t = 10$ when the wave packet is reflected at the ends.

Furthermore, it can be seen that the error increases with time.
A comparison to $\tau = 0.55$ is given for each case.
For $\tau = 0.54$ there is no qualitative difference observable.
By increasing $\tau$, the increase of the errors becomes more prominent, visible in the range of mean errors exceeding those of $\tau = 0.55$.
The range of maximum errors remains almost unchanged, with only a slightly increased lower bound for $\tau = 0.59$.

$\mathcal{E}_\infty$ has values approximately between $1 \times 10^{-2}$ and $1 \times 10^{-1}$, where the upper bound is only reached at the beginning of the simulation.
During free propagation, $\mathcal{E}_\infty$ is closer to $2 \times 10^{-2}$.
At later times - after the second reflection - the dispersed wave packet interacts more with the boundaries.
$\mathcal{E}_\infty$ shows more variations for $\tau = 0.54$ and $\tau = 0.55$ and is increased for $\tau = 0.57$ and $\tau = 0.59$.

The mean error lies between approximately $3 \times 10^{-5}$ and $1 \times 10^{-4}$.
The distance to $\mathcal{E}_\infty$ is again about two orders of magnitude.
After applying the wavelet, the values for $\mathcal{E}_2$ increase over time.

\section{Discussion of the Numerical Results}
\label{sec:results}

The proposed Lattice Boltzmann Method (LBM) successfully simulates fundamental benchmarks for solid mechanics, specifically focusing on shear and uniaxial tension, which represent the basic forms of deformation. These benchmarks are evaluated over a wide range of values for Poisson’s ratio ($\nu$).
They include values close to zero, negative values, and up to the theoretical upper limit, imposed by the Courant-Friedrichs-Lewy (CFL) condition. Despite using a relatively coarse lattice, the grid errors remain low, demonstrating the effectiveness of the method. Although local errors can be considerably higher, the global error provides a reasonable impression of the overall accuracy of the scheme.

The observed deformations align well with mechanical expectations and are comparable to FE reference solutions. The convergence behaviour is as expected: the global maximum error exhibits a first-order convergence rate, indicating the impact of the boundary conditions, while the mean error converges at a second-order rate, reflecting the accuracy of the bulk method itself. This validates the proposed method of including the non-linear source term for a broad range of Poisson’s ratio values.

\subsection{Performance with Boundary Conditions}

For uniaxial tension with inhomogeneous Neumann boundary conditions, the method works well in the interior. However, noticeable issues occur at the corners, where the deformation does not match expectations.
This behaviour at the corners leads to higher errors compared to the shear benchmarks, where the inhomogeneous Dirichlet boundary conditions perform as anticipated without such localized discrepancies.

When comparing constitutive models, the St. Venant-Kirchhoff model performs well under small deformations, and its results agree well with the previous method for linear elasticity.
However, the St. Venant-Kirchhoff model can exhibit instabilities for larger deformations, limiting its applicability. In contrast, the neo-Hooke model remains stable for large deformations.
For the tension benchmark, the neo-Hooke model shows larger deviations due to increased displacements for the same load.
The corner errors observed in the St. Venant-Kirchhoff model persist and are amplified for the neo-Hooke model, especially in scenarios with larger deformations, contributing to the increased errors.

\subsection{Dynamic Simulations}

The method captures dynamic behaviour effectively, with inertia effects clearly visible in the benchmark examples (see Fig.~\ref{fig:num:tension:deform} and~\ref{fig:num:shear:deform}). These effects become particularly prominent in the wave propagation example, where the method successfully simulates the dynamics for the selected Poisson’s ratio. The results demonstrate that longer simulations are feasible, proving the algorithm's potential for dynamic models.

Dynamic scenarios with inhomogeneous Neumann boundary conditions perform as expected, further validating the robustness of the proposed boundary conditions. However, the mean error increases with time, which may eventually lead to simulation instability or failure.

\subsection{Error Analysis and Stability}

Both Neumann and Dirichlet boundary conditions exhibit relatively large errors, especially at the onset of the simulation and during interactions with wave packets. This behaviour is attributed to the boundary conditions being only first-order accurate. Consequently, the finite difference stencil at the boundary also remains first-order, impacting the overall error and the CFL condition.

The choice of the relaxation time $\tau \approx 0.55$ minimizes the growth of error, thus improving stability in the simulations. Deviations from this value result in instability, confirming the suitability of this choice in the benchmark problems.

\section{Summary and Outlook}
\label{sec:summary}

This work proposes a novel method for incorporating non-linear material models into the Lattice Boltzmann (LB) framework.
It represents a direct extension of the previously published method~\cite{faust_dirichlet_2024} for linear elastic materials, both in terms of the underlying algorithm and the handling of boundary conditions.
For the first time, non-linear constitutive behaviour is modelled within the LB framework, where the moment chain approach with a constitutive forcing term serves as the scheme of choice.
Stress and deformation measures are defined in the reference configuration to simplify computations, particularly with regard to topology, as no changes to the underlying lattice are necessary.
Additionally, Neumann- and Dirichlet-type boundary conditions are introduced, maintaining consistency with the treatment used in the linear case.  

The proposed method has been validated by benchmark problems with a range of material parameters.
Two non-linear material models, the St. Venant-Kirchhoff and neo-Hooke models, were considered and compared FE reference solutions.
The results show generally low errors, with quadratic convergence observed for the bulk of the domain and linear convergence at the boundaries.
These findings demonstrate the viability of both the method and the proposed boundary conditions for non-linear solid mechanics.  

The wave propagation example, representing an inherently dynamic simulation, further illustrates the applicability of the method to elastodynamics.
The LBM is inherently well suited for dynamic simulations due to its explicit and time-marching nature. It has the ability to naturally capture wave propagation phenomena.
Although the mean error remains low during longer simulations, it increases with respect to time, and the maximum error is relatively high.
Despite these limitations, the results demonstrate the feasibility of the method for dynamic problems.

However, certain limitations remain.
The first-order accuracy of the boundary conditions contributes to higher overall errors and may ultimately lead to instabilities.
The requirement for first-order finite difference stencils at the boundaries also limits the speed of information propagation, thereby restricting the range of viable material parameters under the CFL condition.  

Improved methods for implementing the boundary conditions are needed, to address the above issues and to enhance both stability and accuracy. However, achieving higher order boundary conditions has proven elusive thus far.
Additionally, the current work employs the BGK collision operator, and identifying the precise stability bounds for this operator lies beyond the scope of this study.
More advanced collision operators have the potential to improve stability and are available in the literature~\cite{coreixas_comprehensive_2019}.  

Further improvements can also be made in the implementation of the method.
A more streamlined evaluation of the material laws, porting the code to high-performance computing frameworks, and a detailed assessment of computational efficiency are promising areas for future work.  

In conclusion, while the method would benefit from advancements in boundary conditions, collision kernels, and finite difference schemes, it demonstrates significant potential as an innovative approach for non-linear elastodynamics.  

\bigskip
\paragraph{Code availability}
For the simulations, the open-source \textit{Python}-package \textsc{SolidLBM} was used.
A snapshot of the code, along with input files for the examples and scripts for data post-processing, is available from Zenodo~\cite{muller_nonlinear_2024}.
The latest version of \textsc{SolidLBM} can be accessed on GitLab~\cite{schluter_pylbm_2024}, distributed under the MIT license.

\paragraph{CRediT}
HM: Data curation, Formal Analysis, Investigation, Methodology, Software, Visualization, Writing – original draft, Writing – review \& editing; EF: Conceptualization, Formal Analysis, Software, Writing – original draft, Writing – review \& editing; AS: Investigation, Methodology, Writing – review \& editing; RM: Conceptualization, Funding acquisition, Project administration, Supervision, Writing – review \& editing

\paragraph{Funding}
This research was funded by the Deutsche Forschungsgemeinschaft (DFG, German Research Foundation)~-- 423809639; 511263698~-~TRR~375

\printbibliography

\end{document}